\documentclass[showpacs,pra,twocolumn]{revtex4}
\usepackage{amsmath}
\usepackage{graphicx}
\usepackage{dcolumn}
\usepackage{bm}

\begin{document}
\title{Quantum Key Distribution with vacuum--one-photon entangled states}
\author{Gian Luca Giorgi}
\email{gianluca.giorgi@roma1.infn.it} \affiliation{INFM Center for
Statistical Mechanics and Complexity and Dipartimento di Fisica,
Universit\`{a} di Roma La Sapienza, Piazzale Aldo Moro 2, I-00185
Roma, Italy}


\begin{abstract}
We present a scheme to realize a quantum key distribution using
vacuum--one-photon entangled states created both from Alice and Bob.
The protocol consists in an exchange of spatial modes between Alice
and Bob and in a recombination which allows one of them to
reconstruct the bit encoded by the counterpart in the phase of the
entangled state. The security of the scheme is analyzed against some
simple kind of attack. The model is shown to reach higher efficiency
with respect to prior schemes using phase encoding methods.
\end{abstract}
\maketitle

Quantum cryptography, or, more correctly, quantum key distribution
(QKD), allows two parties (Alice and Bob) to generate a secret key,
which can be used as a one-time pad, with the guarantee that nobody
else can significantly access the key. Whereas in the case of
classical key distribution the security is connected with the
computational complexity needed to acquire information, QKD secrecy
is based on the laws of quantum mechanics \cite{gisin}. The first
mechanisms exploited is the Heisenberg uncertainty principle,
suggested by Bennett and Brassard in Ref. \cite{bb84} (BB84), while
the other main proposal, introduced by Ekert \cite {ekert}, relies
on the use of entangled states. The BB84 protocol \cite{bb84}\ is
based on the use of two nonorthogonal bases, and has to be
distinguished by other schemes, such as the so-called B92 (Ref.
\cite{b92}) using only two nonorthogonal states, and different
schemes which involve six states \cite{bruss98,bechmann99}. More
recently, a lot of attention has been devoted to the possibility of
using the same criteria of secrecy to realize protocols for
deterministic secure direct quantum communication
\cite{pingpong,marco}. The polarization encoding technique in BB84
is limited by birefringence effects when optical fibers are used as
channels, and thus phase encoding seems to be preferable for
stability in long-distance communication
\cite{rarity,marand,hughes}. More recently, a differential
phase-shift mechanism has been introduced \cite{inoue} which shows
higher efficiency with respect to the previous models.

Here we propose a new scheme of phase encoding based on
vacuum--one-photon entangled states, which involves a complete
symmetry between Alice and Bob, and is designed for stable
transmission. A very different proposal for quantum cryptography
which uses also single-particle entanglement appears in Ref.
\cite{lee03}.
\begin{figure}
\includegraphics[scale=0.75]{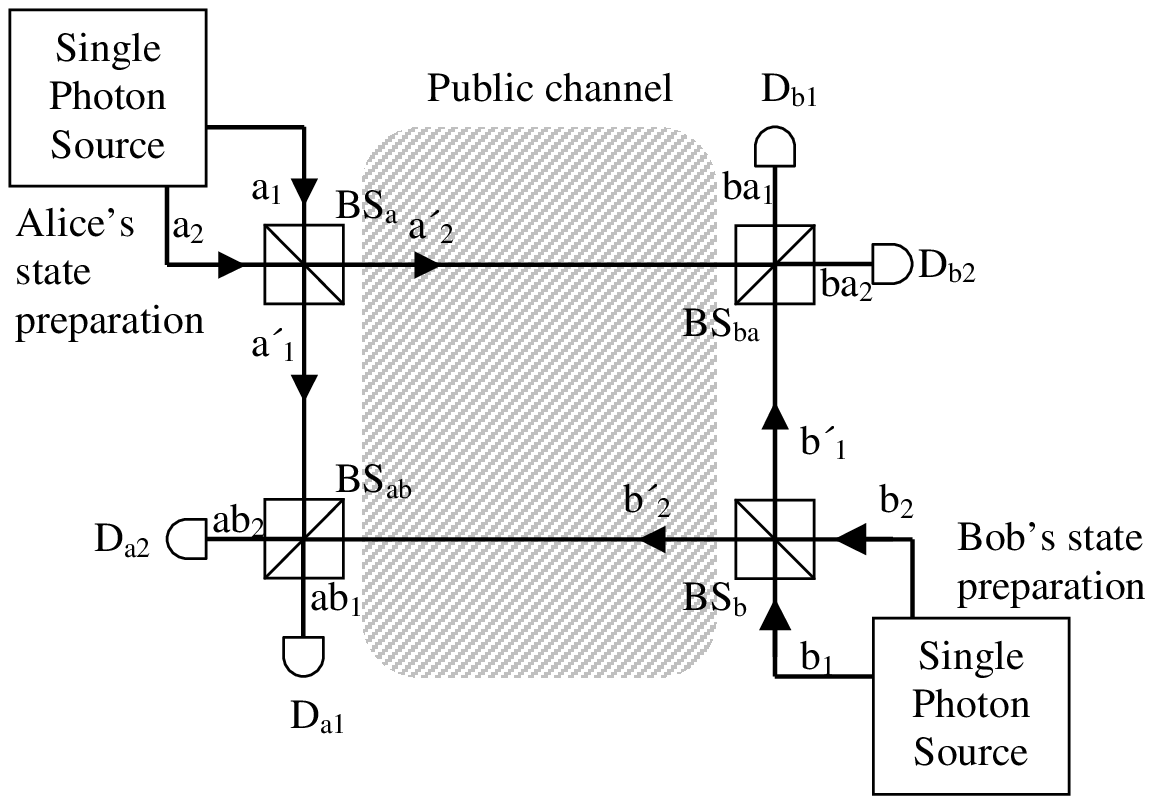}
    \caption{Scheme for QKD using two single-photon
    entangled states. The shaded area represents the public channel
    and is the region where eavesdropping can take place. Alice (left side) and
    Bob (right side) use the respective single-photon sources to
    create two entangled states, encoding the bit on the phase, on the output modes of $BS_a$
    and $BS_b$. Each of them stores one mode in a secure area and
    sends the other mode to the counterpart. The protocol is
    concluded via the recombination on the beam splitters $BS_{ab}$ and
    $BS_{ba}$ and the statement of Alice (the scheme works also exchanging the
    roles) of which detectors ($D_{a1}$ or $D_{a2}$) has counted one
    photon. Comparing this information with his result (click on $D_{b1}$ or $D_{b2}$),
    Bob acquires the secret information.}\label{fig1}
\end{figure}
The scheme is depicted in Fig. 1. Alice wants to create a QKD and to
share it with Bob. She uses a single-photon source which injects the
photon either on the mode $a_{1}$ or on the mode $a_{2}$. The modes
$a_{1}$ and $a_{2}$\ are mixed in a beam splitter ($BS_{a}$) and
then the single
photon is entangled on the two output modes $a_{1}^{\prime }$ and $%
a_{2}^{\prime }$. In terms of field operators the $BS_{a}$ action on
the input-output modes is represented by
\begin{equation}
\hat{a}_{1}^{\dagger }=\frac{1}{\sqrt{2}}\left( \hat{a}_{1}^{\prime
\dagger }+\hat{a}_{2}^{\prime \dagger }\right) \label{a1}
\end{equation}
and
\begin{equation}
\hat{a}_{2}^{\dagger }=\frac{1}{\sqrt{2}}\left( \hat{a}_{1}^{\prime
\dagger }-\hat{a}_{2}^{\prime \dagger }\right) \label{a2}.
\end{equation}
($\hat{a}_{i}^{\dagger}$ creates a photon on the mode $a_{i}$). Thus
the output state is $2^{-1/2}\left( \left| 01\right\rangle +\left|
10\right\rangle \right) $ if the photon is put in the mode $a_{1}$
or $ 2^{-1/2}\left( \left| 01\right\rangle -\left| 10\right\rangle
\right) $ if the photon is put in the mode $a_{2}$. These two
possible choices represent the logic values (the bit) which Alice
wants to add in the QKD. Therefore the bit is encoded in the phase
of the entangled state emerging from $BS_{a}$ which we conveniently
rewrite as $2^{-1/2}\left( \left| 01\right\rangle +n\left|
10\right\rangle \right) $ with $n=-1,1$ ($n=1$ will correspond to
the logic value $1$, $n=-1$ will correspond to the logic value $0$).

Bob, being far apart, realizes the same operation through his own
apparatus and creates the state $2^{-1/2}\left( \left|
01\right\rangle +m\left| 10\right\rangle \right) $ (again, $m=-1,1$)
on the modes $b_{1}^{\prime }$ and $b_{2}^{\prime }$. Obviously, $n$
and $m$ are completely uncorrelated.

Afterwards, Alice (Bob) stores the mode $a_{1}^{\prime }$ ($b_{1}^{\prime }$%
) and sends to Bob (Alice) the mode $a_{2}^{\prime }$
($b_{2}^{\prime }$). Each of them has a second beam splitter
($BS_{ab}$ and $BS_{ba}$) which is used to mix the mode previously
stored with the mode received from the counterpart.

The initial state is
\begin{equation}
\left| \phi \right\rangle =\frac{1}{2}\left( \left| 0_{a_{1}^{\prime
}}1_{a_{2}^{\prime }}\right\rangle +n\left| 1_{a_{1}^{\prime
}}0_{a_{2}^{\prime }}\right\rangle \right) \left( \left|
0_{b_{1}^{\prime }}1_{b_{2}^{\prime }}\right\rangle +m\left|
1_{b_{1}^{\prime }}0_{b_{2}^{\prime }}\right\rangle \right).
\end{equation}
Because of the unitary operation associated to $BS_{ab}$ and
$BS_{ba}$, which consists of field mode relations analogous to Eqs.
(\ref{a1}) and (\ref{a2}), the state $\left| \phi \right\rangle $
becomes
\begin{widetext}
\begin{gather}
\left| \phi \right\rangle =\frac{1}{2\sqrt{2}}[m\left( \left|
0_{ab_{1}}0_{ab_{2}}2_{ba_{1}}0_{ba_{2}}\right\rangle -\left|
0_{ab_{1}}0_{ab_{2}}0_{ba_{1}}2_{ba_{2}}\right\rangle \right)
+n\left( \left|
2_{ab_{1}}0_{ab_{2}}0_{ba_{1}}0_{ba_{2}}\right\rangle -\left|
0_{ab_{1}}2_{ab_{2}}0_{ba_{1}}0_{ba_{2}}\right\rangle \right) +  \notag \\
+\left( mn-1\right) \left( \left|
0_{ab_{1}}1_{ab_{2}}1_{ba_{1}}0_{ba_{2}}\right\rangle +\left|
1_{ab_{1}}0_{ab_{2}}0_{ba_{1}}1_{ba_{2}}\right\rangle \right)
+\left( mn+1\right) \left( \left|
0_{ab_{1}}1_{ab_{2}}0_{ba_{1}}1_{ba_{2}}\right\rangle +\left|
1_{ab_{1}}0_{ab_{2}}1_{ba_{1}}0_{ba_{2}}\right\rangle \right) ].
\label{eq}
\end{gather}
\end{widetext}
The protocol provides a measure realized both by Alice and Bob on
the output modes of $BS_{ab}$ and $BS_{ba}$. The scheme works if and
only if one and only one photon is detected by Alice and one and
only one photon is detected by Bob. Thus the terms corresponding to
two photons entering in one beam splitter and zero photons entering
in the other beam splitter do not contribute, fixing to $1/2$ the
efficiency of the model.

Here we note that, in order to observe quantum interference on
$BS_{ab}$ and $ BS_{ba}$, and this is exactly the situation from
which Eq. (\ref{eq}) is derived, the wave packets impinging the
input arms of the beam splitters are required to be completely
indistinguishable \cite{mandel}. To create such a situation the
stored modes have to be opportunely delayed.

Let us suppose that Alice measures one photon on the mode $ab_{1}$.
The state corresponding to this result is
\begin{eqnarray}
\begin{split}
\left| \phi \right\rangle  =&\frac{1}{2}[\left( mn+1\right) \left|
1_{ab_{1}}0_{ab_{2}}1_{ba_{1}}0_{ba_{2}}\right\rangle +\left(
mn-1\right)\\&\times \left|
1_{ab_{1}}0_{ab_{2}}0_{ba_{1}}1_{ba_{2}}\right\rangle]. \end{split}
\end{eqnarray}
As a consequence, Bob will detect his photon on the mode $ba_{1}$ if
$\ m=n$ or on the mode $ba_{2}$ if $\ m=-n$. If Alice had counted
$``1"$ on the mode $ab_{2}$ the role of Bob's detectors would change
with respect to the relation between $m$ and $n$.

Then, Alice sends on the public channel her result to Bob, who,
comparing the two results, is able to identify the value of $n$ to
add to the key. Due to the complete randomness of the output Alice's
result, there is no connection between the information sent on the
public channel and $n$. We assume that Alice and Bob perform the
measurements in time coincidence. The public statement of which
detector has counted one photon can take place after the entire key
has been realized, as usual in QKD schemes, in analogy with basis
reconciliation in the BB84.

Analyzing the scheme, one can state that the bit exchange is
realized via entanglement swapping \cite{pan} from the modes
$a_{1}^{\prime }$, $a_{2}^{\prime }$ and $b_{1}^{\prime }$,
$b_{2}^{\prime }$ to the modes $ab_{1}$, $ab_{2}$ and $ba_{1}$,
$ba_{2}$, as already suggested in the framework of quantum
cryptography \cite{cabello,song}. The scheme described is in some
aspect related to a cryptographic system recently realized
\cite{degio}: also in that system both Alice and Bob create and
exchange the key. The main differences concern the encrypting method
(the polarization of photons) and a time hierarchy between Alice's
and Bob's operations. As we shall later, this aspect will appear
significant in the security of the scheme.

As in any QKD scheme, we need to consider the possibility that an
eavesdropper (Eve) is trying to gain information, or simply to
disturb the transmission in order to create errors in the reception.

Then, a control procedure has to be introduced. The simple idea is
as follows: for a random subset of bits, during the public
discussion, Alice can claim both which detector has recorded the
photon and the value of $n$ encoded, giving to Bob the possibility
to verify that the global state was not affected by external
interactions.

Apart from limitations on QKD arising from experimental
imperfections regarding generation, transmission, and detection of
qubits \cite{brassard00}, we shall focus our attention on some
simple attack strategy by some external eavesdropper.

First we describe the possibility of an attack only aimed to create
errors in the key. If the disturbance consists in the subtraction of
one photon the protocol automatically fails and there are no effects
on the QKD creation. Better, Eve can act modifying the phase of the
photons traveling in the public channel by an amount between $0$ and
$\pi $. In such a circumstance the control procedure is able to
detect the interference: if the phase change is $\pi $ the role of
detector pairs with respect to $m$ and $n$ is completely inverted,
and when Alice announces both the result and $n$, Bob immediately
discovers Eve's action. More significant is the case of phase change
equal to $\pi /2$: now just about in $50\%$ of cases the action
induces an error, and it is possible that when Alice launches the
control routine Bob does not note the introduction of a third part.
However, after $\nu $ control steps, the probability that Eve is not
revealed is $\left( 1/2\right) ^{\nu }$ and can be arbitrarily
reduced. In the case of phase variation less than $\pi /2$ the
number of control routines to get a given confidence level
increases, but the probability that Eve's action influences the key
decreases.

Let us consider the case that Eve wants actually to get the key.
Since the secret is encoded in the phase of an entangled state, and
one of the components of the state is not accessible to anyone but
Alice, there is no way to get information acting only on the public
mode. Formally, this feature is expressed stating that the reduced
density matrix of a single mode is diagonal and corresponds to a
one-qubit maximally mixed state. The simplest method Eve can use is
the intercept/resend strategy using the same setup as Bob.
Naturally, Eve does not know either $n$ nor $m$ and has to create a
different one-photon entangled state $2^{-1/2}\left( \left|
01\right\rangle +p\left| 10\right\rangle \right) $ ($p=-1,1$), to
mix her state with Alice's state and to wait for Alice's
announcement about the measurement result to conclude the operation.
As in the regular procedure between Alice and Bob, the scheme fails
in half the number of cases, while in the remaining cases Eve
acquires the bit. The quantum bit error rate (QBER) introduced by
Eve in the sifted key (here represented by all bit exchanges with
one photon detected by Alice and one photon detected by Bob) is
$1/2$, due to lack of correlation between $n$ and $p$, while the
amount of information gained by Eve is $1/2$ per bit. Thus,
comparing our model with the BB84, we conclude that, while Eve gets
the same amount of information, she induces a QBER which is twice,
and this feature strongly improves the robustness of the system
against these attacks.

On the other hand, even when the eavesdropping action is performed,
Bob needs to receive a mode from Alice. This aspect involves the
resending strategy that Eve can choose. Eve used one photon to copy
Bob's operation, and whichever is the number of photons sent to Bob
($0,1$, a combination of $0$ and $1$) the total number of photons
revealed by Alice and Bob is no longer $2$, but depends on the
measurement process. Hence, by checking the numbers of contemporary
clicks, Alice and Bob discover the presence of an eavesdropping
action and abort the transmission. Moreover, even if the total
photon number is $2$, by the control routine mentioned above, Eve
can be detected, due to the complete absence of correlation between
$n,m$, and $p$. One can argue that the
eavesdropper can first find $n$ and then send to Bob the correct state $%
2^{-1/2}\left( \left| 01\right\rangle +n\left| 10\right\rangle
\right) $, but Alice's announcement happens after Bob's measurement,
so that the use of coincidence measurements guarantees against this
kind of action.

A more detailed analysis of eavesdropping influence on the counting
rate can be formulated as follows. At the time of her own
measurement, Eve learns how many photons Alice will count. Let us
suppose that she is able (Eve is a quantum devil) first to perform
the measurement and successively choose the resending strategy. The
following situations are possible: (i) Eve knows that Alice will
measure two photons: in such a case the best choice she can make is
to send nothing to Bob; (ii) Alice measures zero photons: now the
choice to minimize the error is to send one photon to Bob; (iii)
Alice measures one photon: now the resending strategy does not
matter. As a result, eavesdropping modifies the number of detected
photons in half of the cases.

Therefore the control about the counting rate represents a powerful
method to reveal eavesdropping to add to the control routine.
Actually, in order to exploit this feature, a multiphoton resolution
is needed, and this not yet fully available in the present
laboratory technology, although some important step has been made
\cite{haderka,franson}.

Naturally, Eve can use an alternative strategy. She can create in
any circumstance two entangled states to share with Alice and Bob,
and, moreover, she can prepare other fake photons to send in order
to enforce both Alice and Bob to count one photon. The cost to pay
for this strategy is the following: due the probabilistic nature of
projections, Alice and Bob expect to measure one photon just in
$1/2$ of cases; then Eve should simulate such behavior leaking a big
amount of information. Thus this strategy is not convenient.

Another simple eavesdropping strategy is the so-called
beam-splitting attack. Let us suppose that a coherent, weak source
of photons is used instead of a single-photon source. Then, with a
probability small but finite, the source can inject two (or more)
photons. In BB84 schemes, the two photons contain the same
information. Then, Eve can subtract one of them and, after the
public discussion, perform the measurement selecting the right
basis. In such a way she acquires the bit without introducing any
kind of noise. Let us analyze what happens in our case, when, for
example, Alice injects two photons onto $BS_a$. The initial state is
\begin{eqnarray}
\begin{split}
\left| \phi \right\rangle =&\frac{1}{2\sqrt{2}}\left( \left|
0_{a_{1}^{\prime }}2_{a_{2}^{\prime }}\right\rangle+\left|
2_{a_{1}^{\prime }}0_{a_{2}^{\prime }}\right\rangle +n\left|
1_{a_{1}^{\prime }}1_{a_{2}^{\prime }}\right\rangle \right)( \left|
0_{b_{1}^{\prime }}1_{b_{2}^{\prime }}\right\rangle
\\&+m\left| 1_{b_{1}^{\prime }}0_{b_{2}^{\prime
}}\right\rangle).
\end{split}
\end{eqnarray}
A simple observation to make is that Eve should be able to factorize
the state $\left( \left| 0_{a_{1}^{\prime }}1_{a_{2}^{\prime
}}\right\rangle +n\left| 1_{a_{1}^{\prime }}0_{a_{2}^{\prime
}}\right\rangle \right) \left( \left| 0_{a_{1}^{\prime
}}1_{a_{2}^{\prime }}\right\rangle +n\left| 1_{a_{1}^{\prime
}}0_{a_{2}^{\prime }}\right\rangle \right)$ from $\left( \left|
0_{a_{1}^{\prime }}2_{a_{2}^{\prime }}\right\rangle +\left|
2_{a_{1}^{\prime }}0_{a_{2}^{\prime }}\right\rangle +n\left|
1_{a_{1}^{\prime }}1_{a_{2}^{\prime }}\right\rangle \right) $, and
to keep one copy. The global nonlocality and the inaccessibility of
the mode $a_{1}^{\prime }$ forbid this kind of eavesdropping
strategy. Obviously, also the protocol fails due to the number of
photons. What matters is that the security of the scheme is robust
with respect to that situation.

Let us come back to analyze the differences between our proposal and
the QKD realized by Degiovanni {\it et al.} \cite{degio}. In that
case there is a time ordering between the encoding operations of
sender and receiver: that is, Alice create a secrete state, Bob acts
on that state, and then resends it to Alice. Therefore an
eavesdropper can extract some information by monitoring the state
before and after Bob's action. In our case we assume that Alice and
Bob perform all operations in coincidence. Therefore all the
information traveling on the public channel is not useful.

On the other hand, the presence of two senders and two receivers
makes our scheme vulnerable versus a subtle strategy: Eve can
short-circuit both Alice and Bob creating two Mach-Zehnder
interferometers. In such a case the two speakers are separated and
each single measurement result depends only, in a deterministic way,
by the initial state created by the respective speaker. Thus Eve has
only to wait for the public communication to perfectly eavesdrop the
bit without introducing noise. Against this kind of attack, we are
helped by the control method introduced by Degiovanni {\it et al}.
Actually, checking the correlation between, for instance, the mode
which Alice stores and the mode which she send to Bob, it is
possible to reveal Eve's presence in half of cases.

The theoretical efficiency $E$ of the scheme can be evaluated
following the criteria introduced in Ref. \cite{holevo}:
\begin{equation}
E=\frac{b_s}{q_t+b_t},
\end{equation}
where $q_t$ is the number of quantum bit exchanged, $b_t$ is the
number of classical bit exchanged, and $b_s$ is the number of secret
bits added to the key. In our case, considering the `` single shot"
efficiency, and the fact that both Bob and Alice add one bit, one
finds $q_t=2$, $b_t=1$, and $b_s=1$, from which follows $E=1/3$. If
the same criterion is applied to Ref. \cite{inoue}, avoiding the use
of active switches, that are not suitable for long distance fiber
communication, we get $E=1/6$. In the case of BB84 protocols the
maximum efficiency that can be reached is $E=1/4$. Thus our proposal
seems to give some advantage. Actually, one should consider some
unavoidable effect that could lower the practical efficiency of the
scheme. For instance, our proposal requires the contemporary
detection of two photons. Thus, the success probability scales with
the square of detection efficiency, in contrast with the usual
situation, where just one detection is needed.

To summarize, we have introduced a method to create a random QKD
based on a mechanism of bit exchange between sender and receiver.
The secret is encoded in the phase of a single-photon entangled
state. Although the encoding is realized only through two orthogonal
states, as in the Goldenberg-Vaidman protocol \cite{gv}, quantum
mechanics guarantees that no information is extracted acting just on
a subsystem, and only the product between Alice's and Bob's states
allows us to extract the key element. The security of the scheme
against simple eavesdropping techniques, such as
intercepting/resending strategy and beam-splitting attack, has been
analyzed. Finally, a comparison with other phase encoding based
schemes has been performed, showing the advantages of our proposal
if addressed to long-distance optical fiber transmission. The scheme
is completely symmetric with respect to the role of Alice and Bob,
and is suitable for information exchange in a sort of quantum
dialogue. Probably, the main obstacle towards a possible realization
of the proposed protocol is represented by the difficulty to achieve
photon number resolution.

The author gratefully acknowledges F. de Pasquale and S. Paganelli.
Special thanks are due to M. Lucamarini for enlightening
discussions.

\end{document}